\documentclass{article}
\usepackage{amsmath,accents}
\usepackage{url}
\usepackage{eufrak}
\usepackage{amsfonts}
\usepackage{amssymb}
\usepackage{amsthm}
\usepackage[figtopcap]{subfigure}
\usepackage[colorlinks,citecolor=blue,urlcolor=blue,linkcolor=blue]{hyperref}

\newcommand\be{\begin{equation}}
\newcommand\ee{\end{equation}}
\newcommand\bea{\begin{eqnarray}}
\newcommand\eea{\end{eqnarray}}
\newcommand\del{\partial}

\usepackage[sort&compress,numbers]{natbib}
\usepackage{tikz}
\usepackage{tikzsymbols}
\usepackage[utf8]{inputenc} 
\usepackage{times}
\usepackage{mathtools}
\usepackage{footnote}
\makesavenoteenv{table}
\usetikzlibrary{arrows,shapes}
\usetikzlibrary{matrix}
\usetikzlibrary{shadows}
\usetikzlibrary{positioning}

\tikzset{
    >=stealth',
    pil/.style={
           ->,
           thick,
           shorten <=2pt,
           shorten >=2pt,}
}

\graphicspath{{./Figs/}}
\begin{document}

\title{\bf Revisiting diagonal tetrads:\\ New Black Hole solutions in f(T) gravity}

\author{Adel Awad${}^{1,2}$, Alexey Golovnev${}^{2}$, Mar\'ia-Jos\'e Guzm\'an${}^{3}$, and Waleed El Hanafy${}^{2}$\\ \\
{\small ${}^{1}${\it 
Department of Physics, Faculty of Science, Ain Shams University, }}\\ 
{\small\it Cairo 11566,
Egypt}\\
{\small ${}^{2}${\it Centre for Theoretical Physics, The British University in Egypt,}}\\ 
{\small\it P.O. Box 43, El Sherouk City, Cairo 11837, Egypt}\\
{\small ${}^{3}${\it Laboratory of Theoretical Physics, Institute of Physics, University of Tartu,}}\\ 
{\small\it W. Ostwaldi 1, Tartu 50411, Estonia}\\ \\
{\small awad.adel@gmail.com}\\
{\small agolovnev@yandex.ru}\\
{\small maria.j.guzman.m@gmail.com}\\
{\small Waleed.Elhanafy@bue.edu.eg}
}
\date{}

\maketitle

\begin{abstract}

We study various forms of diagonal tetrads that accommodate Black Hole solutions in $f(T)$ gravity with certain symmetries. As is well-known, vacuum spherically symmetric diagonal tetrads lead to rather boring cases of constant torsion scalars. We extend this statement to other possible horizon topologies, namely, spherical, hyperbolic and planar horizons. All such cases are forced to have constant torsion scalars to satisfy the anti-symmetric part of the field equations. We give a full classification of possible vacuum static solutions of this sort. Furthermore, we discuss addition of time-dependence in all the above cases. We also show that if all the components of a diagonal tetrad depend only on one coordinate, then the anti-symmetric part of the field equations is automatically satisfied. This result applies to the flat horizon case with Cartesian coordinates. For solutions with a planar symmetry (or a flat horizon), one can naturally use Cartesian coordinates on the horizon. In this case, we show that the presence of matter is required for existence of non-trivial solutions. This is a novel and very interesting feature of these constructions. We present two new exact solutions, the first is a magnetic Black Hole which is the magnetic dual of a known electrically charged Black Hole in literature. The second is a dyonic Black Hole with electric and magnetic charges. We present some features of these Black holes, namely, extremality conditions, mass, behavior of torsion and curvature scalars near the singularity.

\end{abstract}

\section{Introduction}

The motivations for studying modifications of general relativity (GR) are multiple, including the well-known phenomenological problems of cosmology as well as theoretical issues of singularities and quantisation of gravity. One particular branch of this research teeming with alternative candidates corresponds to the geometric foundations of gravity, that is dropping the assumption of the Levi-Civita connection. An interesting subset of such modifications consists of theories based on the teleparallel framework, where the spacetime connection is flat, but allowed to have torsion and/or non-metricity. In the particular case when only torsion is nonvanishing, we retrieve the well-known teleparallel equivalent of general relativity (TEGR), an equivalent version of GR with the same equations of motion \cite{Aldrovandi2013,Golovnev:2018red}, and its modifications such as $f(T)$ which are being intensively studied now \cite{Ferraro:2006jd,Bengochea:2008gz,Ferraro:2008ey,Awad:2017yod,Bahamonde:2021gfp}. 

The usual approach to teleparallel theories requires working with a tetrad instead of just the metric. One can also write the standard GR in this way, and then it has an additional symmetry group of local Lorentz transformations of the tetrad field, the new dynamical variable. At the same time, the action of TEGR preserves this invariance only up to a surface term which is omitted, and at this cost it gets rid of the second derivatives in the Lagrangian. One of its simplest generalisations uses a non-linear function of the torsion scalar $T$ which was in the action of TEGR, and is therefore dubbed $f(T)$ gravity, in analogy with the famous $f(R)$ gravities. The mentioned above difference from GR leads to second order equations of motion, unlike in metric $f(R)$, though with a much more complicated dynamics which has been extensively studied in the recent literature, even though with not too much clarity yet.

Despite all the known foundational problems of these theories \cite{Golovnev:2020zpv}, their cosmological applications receive lots of attention nowadays. At the same time, another interesting and important area of applications corresponds to the search for astrophysical solutions, and unfortunately it is less studied, and also with much less success. Moreover, to the best of our knowledge, these studies were almost fully restricted to static spherically symmetric solutions \cite{Ruggiero:2015oka,DeBenedictis:2016aze,Flathmann:2019khc,Bahamonde:2020vpb,Bohmer:2019vff,Pfeifer:2021njm,Golovnev:2021htv}. We would like to mention that, in fact, other topologies of horizons of Black Hole (BH) type solutions might also be possible, which will be one of the main topics of the present paper.

Of course, the search for spherically symmetric solutions is anyway a reasonable first step towards realistic astrophysical compact objects, and is therefore of utmost importance for testing alternative gravity models. In $f(T)$ gravity this search was rather complicated throughout the time, due to the fact that the dynamical variable is the tetrad instead of the metric, and the theory possesses extra degrees of freedom that produce nontrivial dynamics on top of the usual metric geometry \cite{Li:2011rn,Ferraro:2018tpu,Blagojevic:2020dyq} . It means that only very special cases of tetrads among those which reproduce a given metric are solutions of the corresponding equations of motion, unfortunately with the simplest choices often not being any good.

After adding the spin connection into the game, like in the covariant version of the theory \cite{Krssak:2015oua,Golovnev:2017dox}, any possible tetrad can be used, but the problem translates then to determination of an adequate spin connection for the tetrad preferred. In any case, the enhanced Lorentz symmetry of TEGR is lost, and it equally alters the dynamics in both versions of the theory, as has been proven before \cite{Golovnev:2021omn}. Therefore, in this paper we use the simpler, pure tetrad version of $f(T)$ gravity, or equivalently, the zero spin connection gauge for those who prefer the covariant approach.

Solutions of $f(T)$ gravity equations of motion do qualitatively depend on behaviour of the torsion scalar. Several vacuum solutions with constant torsion scalar have been discussed in the past \cite{Ferraro:2011ks,Nashed:2014sea,Nashed:uja,Bejarano:2014bca,Paliathanasis:2014iva,Nashed:2016tbj}. Such solutions coincide with the usual GR ones (unless with the vanishing derivative of the function, $f_T=0$, which gives a pathology of switched-off gravity), possibly with a cosmological constant, and therefore are not very interesting, unless for studying perturbations around them. With non-constant torsion scalar, the equations become very complicated, even in the case of static ans{\" a}tze with spherical symmetry, and in the spherical coordinates they require non-diagonal tetrads.

Usually, the spherically symmetric solutions are found in terms of series expansions. At first glance, they are not very plausible to be found since there are three equations for two arbitrary functions. However, as it has already been shown, there is a differential relation between them, the generalised Bianchi identity \cite{Golovnev:2020las}. Nevertheless, exact non-GR solutions are still elusive, with probably the only explicit spherically symmetric example employing an everywhere complex tetrad \cite{Bahamonde:2021srr} which is of course highly problematic, even though the metric and the torsion scalar are real. 

Our aim in this paper is to improve the knowledge about spherically symmetric tetrads, and to generalise it to other possible horizon topologies. We will show that in the case of flat, or axially symmetric, horizons there is a natural choice of coordinates in which the antisymmetric equations of motion are automatically satisfied by the diagonal tetrad. In case of magnetic and dyonic BHs, it allowed us to find exact non-trivial solutions, with a non-constant torsion scalar. Electrically charged solutions of this sort were previously found in Ref. \cite{Awad:2017tyz}.

The paper is organised as follows. In Section \ref{Sec:f(T)_gravity} we briefly introduce the theoretical framework of teleparallel gravity and $f(T)$ gravity. In Section \ref{sec:current} we summarise what is know about the tetrad choice for spherically symmetric solutions, including the subsection \ref{sec:Static_Solns} about new static ans{\"a}tze with spherical, axial and hyperbolic symmetry, while in Section \ref{sec:Time-dep_Sol} we extend this  analysis to time-dependence in the tetrad. In Section \ref{sec:flatcart} we present the new choice of the tetrad for the case of axial symmetry which allowed us to find new exact non-constant-$T$ solutions, magnetic in the Section \ref{Sec:magnetic_static} and dyonic in the Section \ref{Sec:dyonic_static}. We present our conclusions in Section \ref{sec:conclusions}.

\section{Teleparallel geometry and $f(T)$ gravity}
\label{Sec:f(T)_gravity}

In this section we briefly describe the teleparallel space which we refer to as a pair ($M$, $e_a$) where $e_a\, (a=1, \cdots, n)$ are $n$ independent vector fields defined on an $n$ dimensional smooth manifold $M$. Let $e_a{^\mu}\, (\mu=1, \cdots, n)$ be their components with respect to a coordinate basis $x^\mu$. This defines $n$ independent 1-forms $e^a=e^a{_\mu} dx^\mu$ such that\footnote{Both the Greek (coordinate) and the Latin (Lorentz) indices are subject to the Einstein summation convention.}
\begin{equation}\label{eq:vielbein}
    e_a{^\mu}e^a{_\nu}=\delta^\mu_\nu, \qquad  e_a{^\mu}e^b{_\mu}=\delta^b_a
\end{equation}
where $\delta$ is the Kronecker delta tensor. On the teleparallel space ($M$, $e_a$) one can define Weitzenb\"{o}ck linear connection,
\begin{equation}\label{eq:W_connection}
    \Gamma{^\alpha}_{\mu \nu}:=e{_a}{^\alpha} \,\partial_\mu e{^a}{_\nu},
\end{equation}
with respect to which the vector fields fulfill the relation  
\begin{equation}\label{eq:}
    \nabla_\mu e^a{_\nu}:=\partial_\mu e^a{_\nu}-\Gamma{^\alpha}_{\mu \nu} e^a{_\alpha}=0 
\end{equation}
where $\nabla$ is the covariant derivative associated with the Weitzenb\"{o}ck connection, and the Latin index in $e^a_{\mu}$ taken as simply numbering the parallelly transported 1-forms, therefore with no connection coefficient associated to it. 

In this sense, the above constraint is called the teleparallelism condition. We write the non-commutation of covariant derivatives acting on arbitrary vector fields $V_{a}$ as
\begin{equation}\label{eq:comm_rel}
\nabla_{\nu}\nabla_{\mu}V_{a}{^{\alpha}} - \nabla_{\mu}\nabla_{\nu}V_{a}{^{\alpha}} = R^{\alpha}{_{\epsilon\mu\nu}}
V_{a}{^{\epsilon}} + T^{\epsilon}{_{\mu\nu}} \nabla_{\epsilon} V_{a}{^{\alpha}}.    
\end{equation}
It can be easily shown that the curvature tensor of Weitzenb\"{o}ck linear connection vanishes identically $R^{\alpha}{_{\epsilon\mu\nu}}\equiv 0$ and the associated non vanishing torsion tensor is given by
\begin{equation}\label{eq:Torsion}
    T{^\alpha}_{\mu \nu}=\Gamma{^\alpha}_{\mu \nu}-\Gamma{^\alpha}_{\nu \mu}=e{_a}{^\alpha} \left(\partial_\mu e{^a}{_\nu}-\partial_\nu e{^a}{_\mu}\right)
\end{equation}
which is antisymmetric in the last two indices by definition.

In addition, one can define a pseudo-Riemannian metric tensor on $M$ in terms of the $\{e_a\}$ as 
\begin{equation}\label{metric}
g_{\mu \nu} := \eta_{ab}e^{a}{_{\mu}}e^{b}{_{\nu}}
\end{equation}
with inverse metric
\begin{equation}\label{inverse}
g^{\mu \nu} = \eta^{ab}e_{a}{^{\mu}}e_{b}{^{\nu}}
\end{equation}
where $\eta_{ab}$ is the Minkowski metric of the tangent space with signature $(-,+,+,+)$. Then one can obtain Levi-Civita connection associated with the metric as
\begin{equation}\label{Christoffel}
\accentset{\circ}{\Gamma}{^{\alpha}}{_{\mu\nu}}:= \frac{1}{2} g^{\alpha \sigma}\left(\partial_{\nu}g_{\mu \sigma}+\partial_{\mu}g_{\nu \sigma}-\partial_{\sigma}g_{\mu \nu}\right).
\end{equation}
It is straightforward to show that both Levi-Civita and Weitzenb\"{o}ck linear connections are metric ones, i.e.
\begin{equation}\label{eq:metricity}
    \accentset{\circ}{\nabla}_{\alpha}g_{\mu \nu}=0={\nabla}_{\alpha}g_{\mu \nu},
\end{equation}
where $\accentset{\circ}{\nabla}$ is the covariant derivative associated with the Levi-Civita connection. 

In the context of the teleparallel geometry one can define an invariant, namely the torsion scalar,
\begin{equation}\label{eq:Torsion_scalar}
    T=T{^\alpha}_{\mu \nu}\, S{_\alpha}^{\mu \nu},
\end{equation}
which differs from Ricci scalar $\accentset{\circ}{R}$, the double contraction of the curvature tensor of Levi-Civita connection, by a total derivative term, where
\begin{equation}\label{eq:Super_potential}
    S{_\alpha}^{\mu \nu}=\frac{1}{4}\left(T{_\alpha}^{\mu \nu}+T^{\nu \mu}{_\alpha}-T^{\mu \nu}{_\alpha}\right)+\frac{1}{2}\left(\delta^\mu_\alpha \, T^{\sigma \nu}{_\sigma}-\delta^\nu_\alpha \, T^{\sigma \mu}{_\sigma}\right).
\end{equation}
This tensor is commonly called the superpotential. It is anti-symmetric in the last pair of indices. Of course, the teleparallel torsion scalar can be used instead of the usual scalar curvature in the Hilbert-Einstein action giving the field equations equivalent to those of the General Relativity (GR) theory, so that we get the Teleparallel Equivalent of General Relativity (TEGR).

One of the simplest generalisations of TEGR is known as $f(T)$ gravity. Its action is given  by
\begin{equation}\label{eq:action}
    \mathcal{S}=-\frac{1}{2\kappa}\int d^4 x \left[e f(T)+\mathcal{L}_m\right].
\end{equation}
where the constant factor is related to the Newton's gravitational constant $G$ by $\kappa=8\pi G$, and $e$ stands for the tetrad $e^{a}_{\mu}$ determinant. The variation of this action \eqref{eq:action} with respect to the tetrad yields the following field equation: 
\begin{equation}\label{eq:field_eqn}
    f_T\mathfrak{G}_{\mu \nu}-\frac{1}{2} g_{\mu \nu}\left(T\, f_T - f\right)+2 S_{\mu \nu}{^\sigma}\, \partial_\sigma f_T=\kappa \mathcal{T}_{\mu \nu}
\end{equation}
where $\mathfrak{G}_{\mu \nu}$ is Einstein tensor, $\mathcal{T}_{\mu \nu}$ is the matter energy-stress tensor, and $f_T\equiv \frac{df(T)}{dT}$. It is to be noted that the TEGR theory is invariant under local Lorentz transformation at the level of the field equations, while this feature is not inherited by the $f(T)$ extension \cite{Li:2010cg}. 

This equation can be rewritten in other ways. For example, one can divide it by $f_T$ to see the change in the effective gravitational constant. On the other hand, for example in cosmology one can go for effective energy density and pressure in modified Friedman equations by rewriting it as
\begin{equation}
    \mathfrak{G}_{\mu \nu}=\kappa\left(\mathcal{T}_{\mu \nu}+\mathfrak{T}_{\mu \nu}\right)
\end{equation}
with 
\begin{equation}
    \mathfrak{T}_{\mu \nu}=\frac{1}{\kappa}\left[\frac{1}{2} g_{\mu \nu}\left(T\, f_T - f\right)-2 S_{\mu \nu}{^\sigma}\, \partial_\sigma f_T + (1-f_T) \mathfrak{G}_{\mu \nu}\right].
\end{equation}
The $\mathfrak{T}_{\mu \nu}$ tensor then represents the modifications of Einstein's field equations due to the $f(T)$ type of action.

As, unlike all the other terms, the term with the superpotential $2 S_{\mu \nu}{^\sigma}\partial_\sigma f_T$ is clearly non-symmetric, the field equations \eqref{eq:field_eqn}
\begin{equation}
{\mathfrak L}_{\mu\nu}=\kappa \mathcal{T}_{\mu \nu}    
\end{equation}
can be split into symmetric and antisymmetric parts with the latter being 
\begin{equation}
 {\mathfrak A}_{\mu\nu}\equiv\frac12 ({\mathfrak L}_{\mu\nu}-{\mathfrak L}_{\nu\mu})=(S_{\mu \nu}{^\sigma}-S_{\nu \mu}{^\sigma})\, \partial_\sigma f_T=(S_{\mu \nu}{^\sigma}-S_{\nu \mu}{^\sigma})\, f_{TT}\cdot \partial_\sigma  T=0.
\end{equation}
It is straightforward to verify that, once the antisymmetric equations are satisfied, then the symmetric part of equations satisfies the generalised Bianchi identity, i.e. its Levi-Civita covariant divergence vanishes \cite{Golovnev:2020las}.

For an $f(T)$ gravity solution with $T=T_c$ constant, the equations of motion (\ref{eq:field_eqn}) take the simple form of
\begin{equation}
f_T(T_c)\mathfrak{G}_{\mu \nu}-\frac{1}{2} g_{\mu \nu}\left(T_c\, f_T(T_c) - f(T_c)\right)=\kappa \mathcal{T}_{\mu \nu}.
\label{eomft}
\end{equation}
These equations are Einstein equations with a rescaled Newton constant $\tilde{G} =\frac{ G}{f_T(T_c)}$ and a cosmological constant
\begin{equation}
\Lambda = \dfrac12 \left(T_c\ - \frac{f(T_c)}{f_T(T_c)}\right).
\label{eq:cosm_const}
\end{equation}
Therefore, if $f(T_c)= T_c f_T(T_c)$, then the GR/TEGR solutions having $T=T_c$ are also solutions of $f(T)$ gravity. And in the case of non-vacuum solutions, the Newton constant should be adjusted. More generally, any TEGR solution having constant $T=T_c$ is a solution of $f(T)$ equations of motion with proper values of effective cosmological and Newton constants.

Note that above we assumed that $f_T\neq 0$. Otherwise it would be a pathological case of switched-off gravity. In particular, if there exists such $T_c$ that $f(T_c)=f_T(T_c)=0$, then any geometrical construction with this constant value of $T$ solves the equations of motion 
(\ref{eq:field_eqn}) in vacuum. The fundamental reason is that every possible linear variation of the $f(T)$ action around it obviously vanishes.

\section{The current situation with Black Hole solutions}
\label{sec:current}

The initial quest for BH solutions of $f(T)$ gravity went a wrong way since it neglected the antisymmetric part of the equations of motion. Therefore, it used diagonal tetrads in spherical coordinates which were not solutions of the equations (\ref{eq:field_eqn}) at all, like for example in the paper \cite{Iorio:2012cm}. However, it is possible to construct solutions using quite a sophisticated tetrad of the form
\be 
e^{a}_{\mu} = \left(
\begin{array}{cccc}
 {A}(r) & 0 & 0 & 0 \\
 0 & {B}(r) \sin\vartheta \cos\varphi &   r \cos\vartheta \cos\varphi & -r\sin\vartheta \sin\varphi \\
 0 & {B}(r) \sin\vartheta \sin\varphi &  r \cos\vartheta \sin\varphi &   r \sin\vartheta \cos\varphi \\
 0 & {B}(r) \cos\vartheta & -r \sin\vartheta & 0 \\
\end{array}
\right).
\ee
A review of what is known about that can be found, for example, in the Ref. \cite{Bahamonde:2021srr}. The equations take a complicated form, so that exact solutions are not known beyond TEGR or constant $T$ constructions. The only exception is an exact solution \cite{Bahamonde:2021srr} with a non-real tetrad, i.e. a tetrad with some components being imaginary. Both the metric and the torsion scalar are real; but since the tetrad is a dynamical variable in the modified teleparallel framework, this situation can be taken as problematic, and in any case it requires more work of defining the model, especially if one wants to study the gravitational perturbations around such solutions.


We now come back to a diagonal tetrad choice and aim at figuring out what can be done for novel BH solutions with that. Of course, in the pure tetrad formulation of a theory like $f(T)$, it is some physical restriction on the ansatz, not simply a gauge choice. Obviously, our motivation is related with the equations which are then easier to be solved. In the covariant language, it is also a simplifying assumption of a diagonal tetrad {\it and} zero spin connection, or to put it in another way, a tetrad which is diagonal in the zero spin connection gauge.

In Sec. \ref{Sec:dyonic_static}, we will show that one can find an exact solution with a diagonal tetrad in the case of flat horizons. Unlike the exact solution from the Ref. \cite{Bahamonde:2021srr}, the tetrad will be real everywhere outside the outer horizon (and inside the inner one), or fully real for a naked singularity solution. In between the two horizons, we will have it complex. This is actually the minimal price to pay. Since we are constructing the standard horizons across which the signs of $g_{tt}$ and $g_{rr}$ change, there is no other way to have it with a diagonal (in these coordinates) tetrad but by using an excursion to imaginary numbers.

Our first task is to see whether any non-trivial new solutions are possible with our choice of diagonal tetrads. In this context, people often talk about "good" and "proper" tetrads. A tetrad is called "good" if its ansatz automatically satisfies the antisymmetric part of equations of motion \cite{Tamanini:2012hg}. We must say that it would be hard to give a precise definition to it, as well as that it actually seems quite strange to have a special name for satisfying a particular part of equations. On the other hand, another approach assumes that a procedure of "switching off gravity" could be used to find {\it the} correct tetrad \cite{Krssak:2018ywd} called "proper". It hinges upon an unjustified assumption that it is possible to objectively separate gravity from inertia, and its whole philosophy neglects the new degrees of freedom. At the same time, not surprisingly, the results of using this recipe have already appeared to be not unique \cite{Emtsova:2021ehh} and even not always correct \cite{Bahamonde:2020snl}, in the sense of giving a "proper" tetrad which is not "good", i.e. not a solution at all.

In other words, explicit solutions are very difficult to find, and it is well-known by now that the spherically symmetric solutions of $f(T)\neq$ TEGR theories can not be found in the simple form of
\begin{equation}
\label{trtetr}
e^{a}_{\mu} = \text{diag}\left(A(r),\ B(r),\ r,\ r\sin(\theta) \vphantom{\int}\right),
\end{equation}
due to the anti-symmetric part of equations not being satisfied. Note that as long as we use usual matter sources, i.e. those with symmetric energy-momentum tensor, it cannot be resolved by considering non-vacuum solutions.

A very important point is that all this was done in spherical coordinates. No claim like that can be made irrespective of the chosen coordinates. Let us state it here explicitly: an abstract claim of impossibility of diagonal tetrads for this kind of solutions is an absolutely wrong statement in itself. The theory is diffeomorphism invariant, and the property of a tetrad being diagonal depends on the choice of coordinates. What is shown in the literature is impossibility of diagonal spherically symmetric tetrads in the standard spherical coordinates. At the same time, one can show that diagonal tetrads in Cartesian coordinates are possible for this metric \cite{Golovnev:2021lki}. They are related to the usual non-diagonal tetrads used in most of the current literature on this topic via the corresponding coordinate transformations.

In this paper, we will rewrite the diagonal tetrads in spherical symmetry with a slightly different choice of coordinates, with no change to the results. However, also hyperbolic and axially symmetric (flat horizon) solutions are to be tried. In the case above, the hyperbolic symmetry corresponds to the simple change of $\sin \longrightarrow \sinh$, and therefore, as is almost immediately obvious, it is of no help. But then we will see that the flat case does give interesting solutions of this sort, in coordinates which correspond to the ones above by substituting unity instead of $\sin$ or $\sinh$ of $\theta$.

\subsection{Generalisation to other horizon topologies}
\label{sec:Static_Solns}

Let us rewrite the usual spherically symmetric ansatz for the metric as
\be 
ds^2 = -A(r)dt^2 + B(r) dr^2 + r^2\left[ \dfrac{d\rho^2}{1-k \rho^2 } + \rho^2 d\phi^2 \right]
\ee 
with $k=1$ and  $\rho \in [-1,+1]$. In principle, a change of coordinates can change the conclusions about possibility of using a diagonal tetrad. However, it is not the case now. We have made a very simple transformation of coordinates, $\theta \longrightarrow \rho=\sin(\theta)$ with no change in $t$, $r$ or $\phi$, nor with any dependence on them in the $\rho$. A tetrad is diagonal in the new coordinates if and only if it was diagonal in the old ones. 

Therefore, the diagonal tetrad
\be
e^a_{\mu} = \text{diag}\left(\sqrt{A(r)},\ \sqrt{B(r)},\ \frac{r}{\sqrt{1-k\,\rho^2}},\ r\rho\right)
\ee
still cannot give us any non-constant $T$ solutions. However, it allows us to treat the cases of hyperbolic ($k=-1$) and flat ($k=0$) horizons uniformly with the usual spherical ones. Moreover, let us also do another innocent coordinate change, $r \longrightarrow {\tilde r}(r)$, so that the angular part of the metric has some arbitrary function of the radial coordinate in front of it, instead of the standard $r^2$.

All in all, we consider the metric
\be 
\label{genmet}
ds^2 = -A(r)dt^2 + B(r) dr^2 + h^2(r)\left[ \dfrac{d\rho^2}{1-k \rho^2 } + \rho^2 d\phi^2 \right]
\ee 
with the following general tetrad which can describe a static gravitational solution in $f(T)$ gravity with spherical, planar or hyperbolic symmetry
\begin{equation}
e^a_{\mu} = \text{diag}\left(\sqrt{A(r)},\ \sqrt{B(r)},\ \frac{h(r)}{\sqrt{1-k\,\rho^2}},\ h(r)\rho\right)
\end{equation}
with an arbitrary function $h(r)$. If such solution possesses a horizon, this will be the symmetry of this null surface. The torsion scalar, in this case, is given by
\be
\label{torscalnew}
T=-2 \frac{h'\, (A\,h)' }{ A B h^2} 
\ee
with $h^{\prime}=\frac{dh(r)}{dr}$.

If the function $h(r)$ is supposed to describe our freedom of choosing the radial variable, then we must assume $h^{\prime}\neq 0$. And in this case, we can actually safely put $h=r$ because $T$ is a scalar, and its behaviour cannot be changed by a coordinate transformation. This is indeed true. If we put $r=r(\tilde r)$ in the metric (\ref{genmet}), the functions $A$ and $h$ remain the same, just calculated at another value of their argument, i.e. $A(r)={\tilde A}(\tilde r(r))$, while the function $B$ also gets multiplied by ${{\tilde r}^{\prime}}^2$. The function $B$ is present in the denominator of the formula (\ref{torscalnew}) for $T$. At the same time, each one of the two derivatives in its numerator gets multiplied by ${\tilde r}^{\prime}$. Therefore, the derivatives of the new radial coordinate with respect to the old one cancel each other, and the quantity $T$ is indeed just a scalar.

Having calculated the anti-symmetric part of the field equations, one gets the following single non-vanishing component
\begin{equation}
{\mathfrak A}_{r\rho}= \frac{1}{2\rho}\cdot f_{TT}  T^{\prime} =0
\label{c-torsion}
\end{equation}
 which, if $f_{TT}\neq 0$, can only be resolved by finding a constant torsion scalar
\be
-2 {h'\, (A\,h)' \over A B h^2}= T_c. \ee
From the equations of motion Eq. \eqref{eomft} we see that such solutions of $f(T)$ gravity are inherited from GR/TEGR, possibly with a cosmological constant, unless the rather pathological case of $f_T(T_c)=0$ is encountered. In other words, these tetrads do not allow us to get non-trivial new solutions.

We would like to mention here that the coordinates we use look similar to the standard treatment of spatially curved cosmologies. Therefore, it is also related to the known fact that these cosmologies do require non-diagonal tetrads \cite{Hohmann:2020zre}, while the case of $k=0$ corresponds to the need for non-diagonal tetrads in the spatially flat cosmology, or even Minkowski space, when written in spherical coordinates.

\subsection{A collection of constant $T$ solutions.}

As was already mentioned above, it is often much easier to find solutions with constant $T$ because in this case the equations reduce to the simple GR ones, or even fully trivialise in the pathological case of $f_T(T_c)=0$. Though those solutions are therefore not very interesting in themselves, let us first discuss possible choices of that sort. We rewrite the eq. \eqref{c-torsion} as the condition of constant torsion scalar (\ref{torscalnew}):
\be
-2 (\ln{h} )'\, (\ln(A\, h))'=T_c\, B.\label{I}\ee

Obviously, the case of $T_c=0$ is very elementary. It is either $h^{\prime}=0$ which goes against our initial assumption of having $h(r)=r$ with some choice of the radial coordinate, or it can be $(Ah)^{\prime}=0$. Let us call these cases ${\bf (i) }$ and ${\bf (ii) }$ respectively. Finally, the case ${\bf (iii) }$ will be about $T_c\neq 0$ which immediately implies non-constant $h$. In this case, suppose we have chosen some function $B(r)$, then define another function $\eta(r)$ such that
\begin{equation}
  \eta^{\prime}=\frac{B}{(\ln{h})^{\prime}}.
\end{equation}
With some arbitrary integration constant in the definition of $\eta$, it transforms the equation \eqref{I} into the one
 \be
\left(T_c\, \eta+ 2 \ln(A\, h)\vphantom{\int}\right)^{\prime}=0 \label{II}\ee
which can immediately be solved as $A(r) =C\cdot \frac{\exp{\left(-{T_c \over 2} \eta(r)\right)}}{ h(r)}$. Note that an additive integration constant in the definition of $\eta$ can then be absorbed into the constant $C$. 

We have just described the geometrical constructions of constant $T$. The equations of motion then take the form of Eq. (\ref{eomft}), and in the cases (i) and (ii) of $T=0$ it even simplifies to $f_T(0) {\mathfrak G}_{\mu\nu}+\frac12 f(0) g_{\mu\nu}=\kappa \mathcal{T}_{\mu \nu}$. The only non-trivial part which remains is the usual computation of the Einstein tensor. Let us have a closer look at all these cases in vacuum.

\subsubsection{Case (i)}

This case of $h^{\prime}=0$ is not about any BHs. The spacetime is a direct product of two 2-dimensional spaces. Here we have $T=0$ and constant $h=h_c$, and the field equations read 
(all the functions $f$ and $f_T$ stand for their values at the $T=0$ point)
\bea &&2{\mathfrak L}_t^t=2{\mathfrak L}_r^r=f-2f_T\cdot{k \over h_c^2}=0, \nonumber\\ && 2{\mathfrak L}_{\rho}^{\rho}=2{\mathfrak L}_{\phi}^{\phi}=f+f_T\cdot{2AA''B-A'^2 B -A'A B' \over 2A^2\,B^2}=0.\eea

Let us first assume that $k\neq 0$. Note that for a non-pathological solution we must have $f_T(0)\neq 0$, and in the case (i) it immediately requires also $f(0)\neq 0$ if $k\neq 0$. The value of $h_c$ trivially follows from the temporal or radial equation, and existence of a real solution obviously depends on the signs of $f(0)$, $f_T(0)$, and $k$. Since $f\neq 0$, the angular equation then requires $A^{\prime}\neq 0$. Therefore, it can be easily solved by multiplying it by this non-zero factor and presenting in the shape of $2fA^{\prime}+f_T\left(\frac{{A^{\prime}}^2}{AB}\right)^{\prime}=0$. The answer is
\be h_c=\pm \sqrt{2k\cdot \frac{f_T(0)}{f(0)}}\qquad \mathrm{and} \qquad
 B(r) ={-  {A'}^2(r) \cdot f_T(0) \ \over  2A^2(r)\cdot f(0)-CA(r)}\ee
 with an arbitrary constant $C$.
 
 For $k=0$ and $h(r)=h_c$, we must have $f(0)=0$ and also satisfy $2AA''B-A'^2 B -A'A B'=0$. In case of a non-degenerate metric with $A^{\prime}\neq 0$, it can be rewritten as $\left(\ln B\right)^{\prime}=\left(\ln \frac{{A^{\prime}}^2}{A}\right)^{\prime}$ giving the obvious solution of $B=C\, {{A'}^2 \over A}$ with an arbitrary constant $C$ and an arbitrary function $A(r)$, as can also be found as the $f(0)=0$ case of above. Another available option is to take $A^{\prime}=0$, i.e. a constant $A$ and an arbitrary function $B(r)$. 

In particular, one message we have from the last option above is that one can describe Minkowski space by a diagonal tetrad
\begin{equation}
e^a_{\mu} = \text{diag}\left(1,\ 1,\ 1,\ \rho\right),
\end{equation}
with one of its planes given in polar coordinates:
\begin{equation}
 ds^2=-dt^2+dr^2+d\rho^2+\rho^2 d\phi^2.   
\end{equation}
The same as for the trivial Minkowski tetrad in Cartesian coordinates, this tetrad solves the vacuum equations of motion\footnote{Actually, its components depend on only one of the coordinates, and therefore, as will be shown below, it would solve the antisymmetric equations automatically, even if it was not a case of constant $T$.} if $f(0)=0$. In line of the discussion in the beginning of this Section, this Minkowski tetrad is an example of a tetrad which is "good" but not "proper". On the other hand, here we have different possible tetrads solving the equations of motion and having the same (Minkowski) metric and the same $T=0$, a fact that represents the so-called remnant symmetries \cite{Ferraro:2014owa}. However such solutions are usually not physically equivalent, for they acquire different properties of perturbations  \cite{Golovnev:2020nln}. It should be taken as manifestation of extra degrees of freedom.

\subsubsection{Case (ii)}

In this case, $A(r)=\frac{A_0}{h(r)}$ with an arbitrary constant $A_0$, which has $T=0$ as well. This leads to the following field equations
\bea  &&2{\mathfrak L}_t^t =f+ 2f_T\cdot \frac{ B \, {h'}^2-hh'B'+2Bhh'' -\,kB^2}{h^2B^2}=0 \nonumber\\
&& 2{\mathfrak L}_r^r=f-2f_T\cdot\frac{k}{h^2}=0,\\
&& 2{\mathfrak L}_{\rho}^{\rho}=2{\mathfrak L}_{\phi}^{\phi}=f+f_T\cdot \frac{ B \, {h'}^2-hh'B'+2Bhh''}{2h^2B^2}=0.\nonumber\eea
Note that the equations with mixed (lower and upper) indices do not have $A_0$ constant in them (though ${\mathfrak L}_{tt}$ would get an overall factor of $A_0$). The reason is that this constant can be absorbed by a constant re-scaling of the time variable, while the solution is static.

By subtracting ${\mathfrak L}^r_r$ from ${\mathfrak L}^t_t$, we see that the second term in ${\mathfrak L}^{\rho}_{\rho}={\mathfrak L}^{\phi}_{\phi}$ must be zero. It requires then $f(0)=0$. And applying the ${\mathfrak L}^r_r$ equation again, we see that also $f_T(0)=0$ in the cases of $k\neq 0$. In other words, for a non-flat horizon, only the option of a rather pathological solution with switched-off gravity is left, with no further restriction for functions $B(r)$ and $h(r)$.

However, for $k=0$, one gets only $f(0)=0$ with no restriction on $f_T(0)$ if $B {h^{\prime}}^2-hh^{\prime}B^{\prime}+2Bhh^{\prime\prime}=0$. The latter equality can be written as $(\ln B)^{\prime}=(h(h^{\prime})^2)^{\prime}$ and solved as $B=C\,h\,(h')^2$, with an arbitrary constant $C$, and $h(r)$ being an arbitrary function. By a radial coordinate choice such that $h=r$, $A=\frac{A_0}{r}$ and $B=Cr$, it leads to a known flat-horizon solution in GR/TEGR which obviously possesses a naked singularity. By constant rescaling of the time variable, we can set $A_0=C$, and therefore $A(r)=\frac{1}{B(r)}$. This solution can be obtained from Schwarzschild-AdS with flat horizon in GR/TEGR after sending the cosmological constant to zero, for example, see the solution in \cite{Emparan:1999gf}.

To summarise, in the case (ii) we always face either the switched-off gravity and a very arbitrary solution then, or a naked singularity which is often considered as a serious pathology.

\subsubsection{Case (iii)}

In this case, $T=T_c\neq 0$ and therefore $h^{\prime}\neq 0$ at any point. We can take an initially arbitrary function $\eta (r)$ where 
\be 
A(r) =C\,{\exp\left(-{T_c \over 2}\, \eta(r)\right) \over h(r)},\ee for some arbitrary constant $T_c$. This leads to a torsion scalar \be T={T_c h'\eta' \over h B},\ee therefore, if $T=T_c\neq 0$, we have \be B(r)=\eta^{\prime}(r)\cdot (\ln{h(r)})^{\prime}=\frac{h^{\prime}\eta^{\prime}}{h}.
\ee 

Upon substituting the expressions for $A$ and $B$ into the equations of motion, and massaging the results a bit, the equations can be written as
\bea  &&2{\mathfrak L}_t^t =f-f_T\cdot\left(\frac{2k}{h^2}+2T_c-\frac{T_c h h^{\prime }{\eta^{\prime}}^2- 2h h^{\prime}\eta^{\prime\prime}+2h h^{\prime\prime}\eta^{\prime}+4{h^{\prime}}^2\eta^{\prime}}{h h^{\prime}{\eta^{\prime}}^2}\right)=0, \nonumber\\
&& 2{\mathfrak L}_r^r=f-f_T\cdot\left(\frac{2k}{h^2}+2T_c\right)=0,\\
&& 2{\mathfrak L}_{\rho}^{\rho}=2{\mathfrak L}_{\phi}^{\phi}=f- f_T\cdot\left(2T_c- \frac{T_c h \eta^{\prime}+2h^{\prime}}{8h {h^{\prime}}^2{\eta^{\prime}}^2}\cdot\left(T_c h h^{\prime }{\eta^{\prime}}^2- 2h h^{\prime}\eta^{\prime\prime}+2h h^{\prime\prime}\eta^{\prime}+4{h^{\prime}}^2\eta^{\prime}\right)\right)=0.\nonumber\eea

Combination of ${\mathfrak L}_t^t$ with ${\mathfrak L}_r^r$ shows that either $f_T(T_c)=0=f(T_c)$, i.e. the pathology of switched-off gravity again, or we need to require
$$T_c h h^{\prime }{\eta^{\prime}}^2- 2h h^{\prime}\eta^{\prime\prime}+2h h^{\prime\prime}\eta^{\prime}+4{h^{\prime}}^2\eta^{\prime}=0.$$
It can be easily taken care of by writing it as $\left(\ln\eta^{\prime}\right)^{\prime}= \left(\frac{T_c}{2}\eta + \ln\left(h^2 h^{\prime}\right)\right)^{\prime}$ which means that $\left(e^{-\frac{T_c}{2}\eta}\right)^{\prime}\propto \left(h^3\right)^{\prime}$ where we used the necessary conditions: $h\neq 0$, $h^{\prime}\neq 0$, $\eta^{\prime}\neq 0$.

Even $k\neq 0$ would anyway not help, since the angular equation contradicts then the radial one. However, in case of $k=0$, and if the theory satisfies the condition $f(T_c)=2T_c\cdot f_T(T_c)$ coming from the radial equation, we can solve all the equations by 
\be \eta(r)=-\frac{2}{T_c} \ln\left(C_1 h^3(r)+C_2\right)\ee
with two arbitrary integration constants $C_1$ and $C_2$, of which $C_1\neq 0$ must be required, in order to not make $B$ vanish. Of course, one could set one of the two constants to unity, and add an additive arbitrary constant to $\eta$ instead.

Since, unlike for $B(r)$, the overall factor in $A(r)$ is not determined by the function $\eta(r)$, this result can equivalently be written down in terms of three arbitrary constants
\begin{equation}
A(r)=C_1 h^2(r)+\frac{C_2}{h(r)}\qquad \mathrm{and} \qquad B(r)=C_3 \frac{{h^{\prime}}^2(r)}{A(r)}
\end{equation}
of which the $C_3$ is not of a big interest since it can be set to $1$ by a simple re-scaling of the time variable by a constant, for it obviously changes the value of $A$ without affecting $B$. By choosing $h=r$, one can check that this is nothing but the known GR/TEGR BH solution in asymptotically de Sitter or anti de Sitter space with flat horizon \cite{Emparan:1999gf,Mann:1997iz,Birmingham:1998nr}, after a simple re-scaling of coordinates. And this is the best of what can be done with this kind of tetrads.

\section{Switching the time dependence on}
\label{sec:Time-dep_Sol}

In this section we take the diagonal tetrad case even further by showing that for a general time-dependent solution of this type the vanishing of the antisymmetric part of the field equations still leads to a constant torsion scalar. In other words, this case cannot be improved by allowing time dependence\footnote{In the next section, we will show a successful tetrad configuration for planar horizons, considering it in both static and time-dependent cases.}.

\subsection{The classical tetrad}
For clarity, let us first illustrate the failure in the classical case. If we take the time-dependent generalisation of the standard (unsuccessful) tetrad (\ref{trtetr}) 
\begin{equation}
e^{a}_{\mu} = \text{diag}\left(A(t,r),\ B(t,r),\ r,\ r\sin(\theta)\vphantom{\int} \right)
\end{equation}
which gives the metric
\begin{equation}
    ds^2 = -A^2(t,r) dt^2 + B^2(t,r) dr^2 + r^2 d\Omega^2,
\end{equation}
then the nonvanishing components of the torsion tensor are easily found to be 
\begin{equation*}
T_{ttr} = A\cdot A^{\prime},  \quad
T_{rtr}  = B \cdot {\dot B},\quad
T_{\theta r \theta}  = r, \quad
T_{\phi r\phi}  = r\sin^2(\theta), \quad
T_{\phi \theta \phi}  = r^2 \cos(\theta) \sin(\theta)
\end{equation*}
where $A^{\prime}\equiv \dfrac{\partial A(t,r)}{\partial r}$ and ${\dot B}\equiv \dfrac{\partial B(t,r)}{\partial t} $.
It yields the following nonvanishing components of the superpotential: 
\begin{equation*}
S_{ttr} = -\dfrac{A^2}{r}, \quad
S_{tt\theta} =- \dfrac{A^2 \cot(\theta)}{2}, \quad
S_{rr\theta}  = \dfrac{B^2 \cot(\theta)}{2}, \quad
S_{\theta t \theta}  =- \dfrac{r^2 \dot B}{2B}, \quad
S_{\theta r \theta} =- \dfrac{r\left( A + r  A^{\prime} \right)}{2A},
\end{equation*}
\begin{equation*}
S_{\phi t\phi}  = \sin^2(\theta) S_{\theta t \theta},\quad
S_{\phi r \phi} = \sin^2(\theta) S_{\theta r \theta}. 
\end{equation*}
Note that both the torsion and the superpotential tensor are antisymmetric in the last two indices, therefore we do not explicitly give here those non-zero components which can be obtained from the ones above by simple interchange of these indices.

We readily see from it that such a tetrad is not possible beyond TEGR and/or constant $T$. For the $f_{TT} S^{\mu\nu\alpha}\partial_{\alpha} T$ term in the equations, i.e. the antisymmetric equations ${\mathfrak A}_{\mu\nu}=0$, the problematic components are $S_{trt},\ S_{t \theta t}, S_{r \theta r}$. They give non-symmetric part of equations of motion in $tr,
\ t\theta, \ r\theta$ components. The first two require $\partial_t  T=0$, and the last one -- $\partial_r  T=0$, and therefore the torsion scalar must be constant. Note that if only one combination of these two derivatives was present in the anti-symmetric equations, it would be possible to find a non-trivial solution, possibly at the cost of some specific energy-momentum tensor of matter.

\subsection{Our tetrad of different topologies}
Our general case is not too different from above, even with all the freedom of an arbitrary function $h(t,r)$. To see this, we take the following tetrad:
\begin{equation}
e^a_{\mu} = \text{diag}\left( \sqrt{A(t,r)},\  \sqrt{B(t,r)},\ \frac{h(t,r)}{\sqrt{1-k\,\rho^2}},\ h(t,r)\rho\right).
\end{equation}
This tetrad leads to the metric 
\begin{equation}
dS^2 = -A(t,r)\, dt^2+B(t,r)\, dr^2+h(t,r)^2\,\left[{d\rho^2 \over 1-k\,\rho^2}+ \rho^2d\phi^2\right]
\end{equation}
where the two-dimensional metric between the square brackets is a two-sphere if $k=1$, a hyperboloid if $k=-1$ or a plane (or a cylinder, or a flat torus) if $k=0$.
Calculating the torsion scalar in this case one finds
\be T(t,r)= -{2 \over  A\, B\,h^2} \,\left({h'\,(A\,h)'}-{\dot{h}\,(B\,h\dot{)}}\right).\ee
The anti-symmetric part ${\mathfrak A}_{\mu\nu}=0$ of the field equations leads to 
\be {\mathfrak A}_{ tr} = f_{TT}\, \left( (\ln h\dot{)}\cdot T^{\prime}-(\ln h)^{\prime} \cdot {\dot T} \right)=0,
\ee
\be {\mathfrak A}_{\rho\,t} = {1 \over 2\rho}\cdot f_{TT}  {\dot  T}=0,
\ee
\be 
{\mathfrak A}_{\rho\,r} = {1 \over 2\rho}\cdot f_{TT} T^{\prime}=0
\ee
where $\,A' \equiv \del_r A$ and $\dot{A} \equiv \del_tA$. Since, for the tetrad considered, the most general torsion scalar can only depend on $t$ and $r$, $T=T(t,r)$, then (any two of) the above three field equations lead to a constant torsion scalar, $T=T_c$.

We conclude that, equivalently to the static cases above, time-dependent diagonal tetrads with spherical and hyperbolic symmetries lead to trivial solutions in a generic $f(T)$ theory, i.e., solutions with constant torsion scalar. The same is true of the axial symmetry in these (polar) coordinates. However, this symmetry  naturally allows another choice of coordinates, too.

\section{Flat horizons with Cartesian coordinates}
\label{sec:flatcart}

Now we will focus on the flat horizon ($k=0$) case which can be written in radically different coordinates. The corresponding diagonal tetrad is different from the one we considered in the previous sections. This is similar to the change of coordinates between polar and Cartesian ones. It does not preserve the property of a tetrad to be diagonal. In other words, we are going to study a genuinely different tetrad,
\begin{equation}
\label{flathortet}
e^a_{\mu} = \text{diag}\left(\sqrt{A(r)},\ \sqrt{B(r)},\ h(r),\ h(r)\vphantom{\int}\right)
\end{equation}
which reproduces the metric $$ds^2=-A(r)dt^2+B(r)dr^2+h^2(r)(d\theta^2+d\phi^2).$$
Let us mention that these coordinates are topologically more natural and can cover the whole space, unlike the polar type of coordinates in case of torical or cylindrical topology, as opposed to a whole infinite plane. In the context of teleparallel studies, it also means that this tetrad is globally better defined.

It yields the following torsion scalar
\be 
\label{flatTsc}
T = -\dfrac{2h^{\prime}\left( h A\right)^{\prime} }{h^2 AB },
\ee 
and the components of the three equations of motion are given by
\bea 
2{\mathfrak L}_t^t & = &  f + \dfrac{f_T}{AB^2h^2}\cdot\left(2A^{\prime}h^{\prime}Bh - 2 B^{\prime} h^{\prime} hA + 4 h^{\prime 2}AB + 4 h^{\prime\prime}hAB \right) + \dfrac{4h^{\prime}f_{TT} }{Bh} \cdot T^{\prime} = 0, \\
2{\mathfrak L}_r^r & = &  f +  \dfrac{4h^{\prime}f_T}{ABh^2}\cdot \left(h A^{\prime} + h^{\prime} A\right)= 0, \\
2{\mathfrak L}_{\theta}^{\theta} =2 {\mathfrak L}_{\phi }^{\phi} & = & f + \dfrac{\left(hA^{\prime} + 2Ah^{\prime}\right)f_{TT}}{ABh} \cdot T^{\prime} + \dfrac{f_{T}}{2A^2B^2h^2}\cdot\left(\vphantom{\int} -AhB^{\prime}\left(hA^{\prime}+2Ah^{\prime}\right) \right. \\
& & \left. + B\left(4A^2h^{\prime 2} -h^2(A^{\prime 2} - 2A A^{\prime\prime} ) + 2Ah(3A^{\prime}h^{\prime} + 2Ah^{\prime\prime} ) \vphantom{\frac{a}{a}}\right) \vphantom{\int}\right). 
\eea 
Using this tetrad and these equations, we were able to find exact, and with non-constant torsion scalar, solutions when coupling the $f(T)$ gravity to an electromagnetic field. But to start with, let us mention that of course it is possible to look for a constant $T=T_c$ solution also in this case. After all, the functional shape of the torsion scalar is the same as before. 

For example, for the function $f(T)=T_0+T+\beta \, T^2$ and with the standard choice of the radial coordinate, which we will also use in the next sections, a vacuum solution with constant torsion scalar (of our case (iii) above) can be given by
\bea h(r)=r, \qquad
A(r) = {C_1 \over r} +C_2\,r^2, \qquad
B(r) = \frac{C_3}{A(r)}
\label{constTsol}
\eea
with initially arbitrary constants $C_i$ restricted by the torsion scalar (\ref{flatTsc}) value
\be 
\label{constTsc}
T_c = \dfrac{-1\pm \sqrt{1+12 \beta T_0} }{6\beta}.
\ee 
This value of $T$ is a consequence of the radial equation requiring $f(T_c)=2T_c \cdot f_T(T_c)$ again. And once the torsion scalar is made to be constant, the equations cease to depend on which one of the $k=0$ tetrads was used.

However, in this case the anti-symmetric part of the field equations is vanishing, without any constraint on the three functions in the metric, nor on the value of the torsion scalar or the function $f$ defining the model. There is another reason for which we need a matter source. Indeed, carefully looking at the radial equation, we see that it identically takes the form of $f(T)-2T \cdot f_T(T)=0$ when in vacuum. Generically, it is an algebraic equation demanding some particular constant value of $T$. Since it is the symmetric part of equations, what makes them not solvable in vacuum, then any usual matter source can have its influence on the situation. We will use it to give less trivial solutions in the sections to follow. Our exact solutions there will be in a non-vacuum case, though with a very natural type of matter, that of electromagnetic field. 

Note also that, with the requirement of $f(T)=2T \cdot f_T(T)$, the only chance to obtain non-constant $T$ solutions in vacuum is for the theories with $f(T)\propto\sqrt{\pm T}$. This is an unpleasant situation when a theory allows for only one particular sign of $T$. Also, presence of matter would be allowed only with vanishing radial component of the energy-momentum tensor. Interestingly, it has been noticed before that functions of this type are special in $f(T)$. For example, any standard de Sitter, i.e. $e^a_{\mu}=\mathrm{diag} \left(1, e^{Ht}, e^{Ht}, e^{Ht}\right)$ with constant $H$, is a vacuum solution in this case \cite{Rodrigues:2012qua}. Indeed, this is a constant $T$ construction, and being a maximally symmetric spacetime with ${\mathfrak G}_{\mu\nu}=-\frac{T}{2} \cdot g_{\mu\nu}$, it brings the equations (\ref{eomft}) to the form of $\left(\frac12 f - f_T T\right)\cdot g_{\mu\nu}=\kappa \mathcal{T}_{\mu \nu}$. Only the energy-momentum of a cosmological constant type is allowed then in any $f(T)$ gravity, and then this is an algebraic equation for determining the Hubble constant $H$. So far, this is nothing unusual, of course. However, these models of $\sqrt{T}$-type make the l.h.s. vanish identically, and therefore allow only for fully vacuum solutions, though with absolutely arbitrary $H$.

Before we describe our new solutions, let us discuss a bit more about the general properties of this kind of tetrads.

\subsection{The general theorem}

The case of the tetrad from this Section can be generalised to the following statement which also includes spatially flat, or even Bianchi I, cosmology:

{\bf Theorem.} A diagonal tetrad with all components depending on only one of the coordinates (in which it is diagonal) automatically satisfies the antisymmetric part of $f(T)$ equations of motion.

To prove the theorem, we take a diagonal tetrad with components that depend only on one coordinate $\xi$:
\begin{equation}
e^a_{\mu}=A_{(\mu)}(x_{\xi})\cdot \delta^a_{\mu}.
\end{equation}
In other words, it has four non-zero components equal to four arbitrary functions $A_{(\mu)}$ of one particular coordinate $x_{\xi}$. Obviously, we will never assume summation over $\xi$; and any summation over an index which also has some appearance in brackets will always be indicated explicitly.

The torsion tensor can be written as
\begin{equation}
T^{\alpha}_{\hphantom{\alpha}\mu\nu}=\frac{A^{\prime}_{(\alpha)}}{A_{(\alpha)}}(\delta^{\xi}_{\mu}\delta^{\alpha}_{\nu}-\delta^{\xi}_{\nu}\delta^{\alpha}_{\mu})    
\end{equation}
where $A^{\prime}\equiv\frac{\partial A}{\partial x_{\xi}}$, or
\begin{equation}
T_{\alpha\mu\nu}=A^{\prime}_{(\alpha)}A_{(\alpha)}(\delta^{\xi}_{\mu}\eta_{\alpha\nu}-\delta^{\xi}_{\nu}\eta_{\alpha\mu})    
\end{equation}
upon lowering the index.

The torsion vector is
\begin{equation}
T_{\mu}\equiv T^{\alpha}_{\hphantom{\alpha}\mu\alpha}=\delta^{\xi}_{\mu}\cdot \sum_{\alpha\neq\xi}\frac{A^{\prime}_{(\alpha)}}{A_{(\alpha)}}    
\end{equation}
and has only one non-zero component, $T_{\xi}$.

The contortion is also easily found:
\begin{equation}
K_{\alpha\mu\nu}=\frac12(T_{\alpha\mu\nu}+T_{\nu\alpha\mu}+T_{\mu\alpha\nu})=A^{\prime}_{(\mu)}A_{(\mu)}(\delta^{\xi}_{\alpha}\eta_{\mu\nu}-\delta^{\xi}_{\nu}\eta_{\mu\alpha}) 
\end{equation}
which finally gives the superpotential
\begin{equation}
S_{\alpha\mu\nu}=\frac12 \left(K_{\mu\alpha\nu}+g_{\alpha\mu}T_{\nu}-g_{\alpha\nu}T_{\mu}\right)=\frac12(\eta_{\alpha\mu}\delta^{\xi}_{\nu}-\eta_{\alpha\nu}\delta^{\xi}_{\mu})\cdot A^2_{(\alpha)}\sum_{\beta\neq\xi,\alpha}\frac{A^{\prime}_{(\beta)}}{A_{(\beta)}}.    
\end{equation}
The right factor in the last expression can be easily derived assuming that $\alpha\neq\xi$. However, in case of $\alpha=\xi$ its left factor is anyway zero. And indeed $S_{\xi\mu\nu}=0$, so that this formula can always be used.

In particular, since $T$ can only depend on the $\xi$ coordinate, we see that $S_{\mu\nu \alpha}\partial^{\alpha}  T=S_{\mu\nu \xi}\partial^{\xi}  T$, and it has only symmetric, even diagonal, components which means that this is always a "good tetrad", in the sense of automatically solving the antisymmetric part of equations. The theorem is proven.

This theorem applies to many different cases, for example to the standard spatially flat cosmology. In our case of $A_{(t)}=\sqrt{A(r)},\ A_{(r)}=\sqrt{B(r)},\ A_{(\theta)}=A_{(\phi)}=r$ we get the only non-zero components of $S_{\mu\nu\xi}$ for the $f_{TT}$ term: $S_{ttr}=-\frac{A}{r}$ and $S_{\theta\theta r}=S_{\phi\phi r}=\frac{r^2}{2}(\frac{A^{\prime}}{2A}+\frac{1}{r})$ with the radial coordinate in the role of $\xi$. Finally, for completeness of these calculations, we present the torsion scalar:
\begin{equation}
 T= S_{\alpha\mu\nu}T^{\alpha\mu\nu}=-\frac{\eta^{\xi\xi}}{A^2_{(\xi)}}\cdot \sum_{\alpha\neq\xi}\left(\frac{A^{\prime}_{(\alpha)}}{A_{(\alpha)}}\sum_{\beta\neq\alpha,\xi}\frac{A^{\prime}_{(\beta)}}{A_{(\beta)}}\right)   
\end{equation}
which then gives $ T = -\frac{2}{B r}(\frac{A^{\prime}}{A}+\frac{1}{r})$.

Interestingly enough, if we take $A_{(\theta)}=A_{(\phi)}=h(r)$, then this result is  $ T = -\frac{2 h^{\prime}}{B h}(\frac{A^{\prime}}{A}+\frac{h^{\prime}}{h})$ which coincides with our result of the previous section. The case of $k=0$ in those tetrads gives another tetrad for the same metric we have here. Reproducing the same torsion scalar then can be taken as yet another case of the remnant symmetry \cite{Ferraro:2014owa}. At the same time, it shows that the actual meaning of remnant symmetry is rather limited, especially in a non-constant $T$ scenario. Those two cases of $k=0$ tetrad have the same metric and the same torsion scalar, but one of the cases does solve the antisymmetric equations while another one doesn't.

\subsection{Comment on time dependence}

Notice that, the same as in the static case, the tetrad with a planar symmetry of the form
\begin{equation}
e^a_{\mu} = \text{diag}\left( \sqrt{A(t,r)},\ \sqrt{B(t,r)},\ h(t,r),\ h(t,r)\vphantom{\int}\right),
\end{equation}
can not be obtained by a coordinate transformation from the $k=0$ case of the tetrad from the previous sections, for under the transformation relating the two metrics, that tetrad would get off-diagonal terms. In other words, this is a different tetrad.

With the non-trivial time-dependence, the anti-symmetric part of the field equations becomes non-trivial even for this new tetrad. But it leads to just a single equation in this sector, namely
\be {\mathfrak A}_{ tr} = f_{TT}\, \left( (\ln h\dot{)}\cdot  T^{\prime}-(\ln h)^{\prime} \cdot {\dot T} \right)=0\ee
which generically does not lead to a constant torsion scalar. For example, the equation is satisfied if $h(t,r)\sim (rt)^a$ and $T \sim (rt)^b$, where $a$ and $b$ are some constants. Also, setting $h(t,r)=r$, we need a constant in time torsion scalar but possibly dependent on $r$. This is a certain restriction for the functions in the tetrad, also with an off-diagonal $tr$ component in the symmetric equations, so that the equations might well get too restrictive in vacuum. However, having solved the $\mathfrak A$ part, one can try to look for a solution with matter.

\section{Magnetic static solutions with axial symmetries}
\label{Sec:magnetic_static}

In this section, we present a new magnetically charged solution in Maxwell-$f(T)$ gravity. The equations we use are the $f(T)$ equations (\ref{eq:action}) of motion with electromagnetic field as a source and Maxwell equations with no charges or currents:
\be
    f_T\mathfrak{G}_{\mu \nu}-\frac{1}{2} g_{\mu \nu}\left(T\, f_T - f\right)+2 S_{\mu \nu}{^\sigma}\, \partial_\sigma f_T=\kappa \mathcal{T}_{\mu \nu}, \hspace{0.5 in} \accentset{\circ}{\nabla}_{\nu} F^{\mu\nu}=0, \ee
where the energy-momentum tensor of the electromagnetic field 
\be \mathcal{T}_{\mu \nu}=\frac{1}{4\pi}\left(F_{\mu\alpha}\,F_{\nu}^{\hphantom{\mu}\alpha}-\frac14 F_{\alpha\beta}F^{\alpha\beta}g_{\mu\nu}\right),\ee
is given in terms of its strength tensor $F_{\mu\nu}\equiv \partial_{\mu} {\mathcal A}_{\nu}-\partial_{\nu} {\mathcal A}_{\mu}$. There is also another, a slightly more mathematical way to write the electromagnetic part of the equations by using the Hodge dual of the field strength $H_{\mu\nu}\equiv *F_{\mu\nu}={1 \over 2}\, \epsilon_{\mu\nu\alpha\beta}\,F^{\alpha\beta}$: 
\be \mathcal{T}_{\mu \nu}=\frac{1}{8\pi}\left(F_{\mu\alpha}\,F_{\nu}^{\hphantom{\mu}\alpha}+H_{\mu\alpha}\,H_{\nu}^{\hphantom{\mu}\alpha}\right),\ee
 Notice that the equations of motion are therefore invariant under electric-magnetic duality, or changing the 2-forms as $F\rightarrow H$ and $H\rightarrow -F$. We will comment on this symmetry after presenting new magnetic and dyonic solutions afterwards. 
 
 Note that, of course, in the theoretical physics literature the factors of $4\pi$ are often omitted. In order to have our BH formulae in a more familiar shape, we keep it and also take $\kappa=8\pi G$, so that the r.h.s. of the equations acquires the form of $\kappa {\cal T}_{\mu\nu}= 2G\left(F_{\mu\alpha}\,F_{\nu}^{\hphantom{\mu}\alpha}-\frac14 F_{\alpha\beta}F^{\alpha\beta}g_{\mu\nu}\right)$ with $G$ being the gravitational constant. And for brevity, we put $G=1$. If needed, the explicit factor of $G$ can be reconstructed in all expressions below by multiplying each $q$ and $p$ by $\sqrt{G}$. 

We are interested in solutions with axial symmetry which can be represented by the diagonal tetrad (\ref{flathortet}) in which we chose the standard radial coordinate such that $h(r)=r$. Also, it proves convenient to write $B(r)=\frac{W(r)}{A(r)}$ to identify the asymptotic behaviour of the solution. Therefore, we take the tetrad to be
\begin{equation}\label{eq:static_tetrad}
e^a_{\mu} = \text{diag}\left( \sqrt{A(r)},\ \sqrt{\frac{W(r)}{A(r)}},\ r\ ,r\right),
\end{equation}
in the coordinates $(t,r,\theta,\phi)$, with $\theta$ and $\phi$ not necessarily being angle-like coordinates. The metric is
\be
dS^2 = - A(r) dt^2 + \dfrac{W(r)}{A(r)} dr^2 + r^2 \left( d\theta^2 + d\phi^2 \right).
\ee
Obviously, the usual (anti)-de-Sitter asymptotic behaviour corresponds to $W(r)$ tending to a constant positive value which can be conveniently set to unity.

This is the same geometry that we had in the previous Section. On top of that, we will assume the following form of the gauge potential 
\be {\cal A}= {\cal A}_4(\theta)\, d\phi.\label{eq:gauge_potential}\ee
 The Maxwell equations are easy to solve in this case. Indeed, the only non-vanishing component of the field strength is $F_{\theta\phi}=-F_{\phi\theta}=\partial_{\theta}{\cal A}_4(\theta)$, and since the metric is diagonal and with components depending only on $r$, the only non-trivial equation tells us that
$$0=\frac{1}{\sqrt{-g}}\partial_{\mu}\left(\sqrt{-g}F^{\mu\phi}\right)
=\partial_{\theta}F^{\theta\phi}\propto \partial_{\theta}F_{\theta\phi}=\partial_{\theta}\partial_{\theta}{\cal A}_4(\theta)$$
which, modulo a gauge choice of ${\cal A}_4(0)=0$, gives ${\cal A}_4(\theta)=p\theta$ with the constant $p$ representing the magnetic charge of the BH.

\subsection{GR/TEGR}
\label{Sec:magnetic_static_TEGR}

Before we present our new magnetic solution, it is instructive to briefly review a known magnetic solution in Maxwell-TEGR. Let us consider the TEGR case, i.e. $f(T)=T_0+T$. Obviously, we should get magnetically charged flat-horizon BHs of the usual GR with a cosmological constant given by $\frac{T_0}{2}$.

For the tetrad \eqref{eq:static_tetrad}, we obtain the components of the superpotential tensor \eqref{eq:Super_potential} 
\be
S_{trt} = - S_{ttr} = \frac{A(r)}{r},
\qquad
S_{\theta\theta r} = S_{\phi\phi r} = - S_{\theta r\theta} = -S_{\phi r\phi} = \dfrac{r}{4A}(rA'+2A )
\ee
and the torsion scalar \eqref{eq:Torsion_scalar} as above (\ref{flatTsc})
\be
T= -\dfrac{2}{r^2 W}( rA' + A).\label{eq:static_Tsc}
\ee
And the only non-zero component of the field strength is $F_{\theta\phi}=-F_{\phi\theta}=\partial_{\theta}{\cal A}_4(\theta)=p$. Therefore, the value of the torsion scalar can be derived from the radial equation 
$\frac12 f - f_T T=\kappa {\cal T}^r_r$. With $f(T)=T_0+T$ and $\kappa{\cal T}^r_r=-\frac12 F_{\mu\nu}F^{\mu\nu}=-\frac{p^2}{r^4}$, we immediately get
\be
T(r)= T_0+ \frac{2p^2}{r^4}.\label{eq:mag_sing}
\ee
The presence of magnetic charge allows us to have a non-constant $T$ solution.

Moreover, with $G=1$, we can easily find the full energy-momentum tensor of the electromagnetic field as
$$\kappa {\cal T}^{\nu}_{\mu}=\mathrm{diag} \left(-\frac{p^2}{r^4}, \, -\frac{p^2}{r^4}, \, \frac{p^2}{r^4},\, \frac{p^2}{r^4} \right).$$
Then the field equations can be solved as 
\be A=C\left(-{T_0 \over 6}\, r^2 \, -{2m \over r} + {p^2 \over r^2}\right) \qquad W= C,  \qquad {\cal A}_4= p\, \theta \ee
where the cosmological constant is given by $\frac{T_0}{2}$, and there are three arbitrary constants, with $p$ being the previously defined magnetic charge and $m$ related to the BH mass. The arbitrary constant $C>0$ is a proportionality coefficient between $B$ and $\frac{1}{A}$, and as such it can be set to $C=1$ by simply re-scaling the time coordinate, and this is a reasonable choice to make in order to have a familiar form of the asymptotic behaviour.

\subsection{Quadratic polynomial $f(T)$ gravity}
\label{Sec:magnetic_static_fT}

Before moving to a magnetic solution for the quadratic polynomial gravity, $$f(T)=T_0+T+\beta \, T^2,$$ we need to set a value for $T_0$ in order to simplify the resulting solutions, for otherwise these solutions will not have a simple analytic form.  Recall that the uncharged solution, for the tetrad ansatz \eqref{eq:static_tetrad}, can be found as (\ref{constTsol})
\be A(r)= C_1\,r^2+C_2/r, \hspace{0.5 in} W(r)=C_3\ee
with a convenient choice of $C_3=1$ easily achievable by a constant time re-scaling. This is irrespective of the particular choice of the function $f$.

A natural fine-tuning is suggested by the value $T_c = \dfrac{-1\pm \sqrt{1+12 \beta T_0} }{6\beta}$ of the torsion scalar (\ref{constTsc}) in the vacuum solutions.
Notice that by choosing
\begin{equation}
  T_0=-{1 \over 12 \, \beta}  
\end{equation}
one gets rid of the square root term in the above expression, and therefore it also simplifies a lot the form of the magnetic solution and the dyonic solution in the coming section. For the uncharged case in this quadratic polynomial $f(T)$ gravity, one then has the constant torsion scalar $T_c=2T_0=-\frac{1}{6\beta}$.

In other words, our model is given by
\be
f(T)=T_0 + T -\frac{T^2}{12T_0} = -\frac{1}{12\beta}+ T + \beta T^2.
\ee
The required torsion scalar can be found in this case by observing that $\frac12 f - f_T T=\frac{1}{2T_0}\left(T_0 - \frac{T}{2}\right)^2$. Since according to the radial equation it must be equal to $-\frac{p^2}{r^4}<0$, we need to demand $T_0<0$, or equivalently $\beta>0$. Clearly the obtained solution has no TEGR limit since $\beta=0$ is not allowed. This case is similar to the charged AdS BH solution in higher-dimensional $f(T)$ gravity \cite{Nashed:2018cth}. After that, the torsion scalar can be found as
\be T = 2T_0\pm 2\sqrt{-2 T_0}{\ |p\ | \over r^2}= -{1 \over 6 \beta}\pm \sqrt{\frac{2}{3\beta}}{\ |p\ | \over r^2}. \label{eq:quad_Tsc_mag}\ee
Note that the two possible solutions come from the case of the quadratic equations for finding $T$. However, we have there $\ |p\ |=\sqrt{p^2}$ which is not surprising since the sign of the charge cannot influence the geometry, with ${\cal T}_{\mu\nu}$ being quadratic in $F_{\mu\nu}$. Below we will omit the plus-minus and the modulus signs, but one should keep in mind that the sign of $p$ can be chosen arbitrarily, irrespective of the real sign of magnetic charge of the BH, and it represents two different possible gravitational solutions.

The full field equations can be solved as 
\be A =C\cdot\left( {r^2 \over 36 \beta}-{2m \over r}+{3p^2 \over 2 r^2}+{\sqrt{6\beta} p^3 \over 3 r^4}\right)= C\cdot\left( -{T_0 r^2 \over 3}-{2m \over r}+{3 p^2 \over 2 r^2}+{ p^3 \over 3\sqrt{-2 T_0}\cdot  r^4}\right), \label{eq:quad_A_mag}\ee
\be W=C\cdot \left(1+{\sqrt{6\beta} p \over r^2}\right)^2 = C\cdot \left(1+{ p \over \sqrt{-2 T_0}\cdot r^2}\right)^2, \hspace{0.3 in} {\cal A}_4= p\, \theta. \label{eq:quad_B_mag}\ee
The constant $C>0$ can again be naturally set to $C=1$ by a proper re-scaling of the time variable.

This solution is the magnetic dual of the solution presented in \cite{Awad:2017tyz} and its rotating version\footnote{The procedure to add angular momentum to 4-dimensioal spacetime has been developed in GR context in \cite{Lemos:1994xp} and generalized to $N$-dimension in \cite{Awad:2002cz}.} in \cite{Awad:2019jur}.
Notably, in the quadratic polynomial $f(T)$ gravity, as $r \to 0$, the teleparallel torsion scalar diverges as $T \to {1 \over r^2}$ , which shows that the quadratic $f(T)$ gravity smooths the singularity in comparison to Maxwell-TEGR case, namely Eq. \eqref{eq:mag_sing}. It is important to comment here on the possibility of generating another solution which in this case an electric solution. I.e., with electric charge $q$ instead of magnetic charge $p$ since the above field equations are invariant under $F\rightarrow H$ and $H\rightarrow -F$ which takes $q\rightarrow p$ and $p\rightarrow -q$. This produces the previously known solution \cite{Awad:2017tyz}. Therefore, no new solution is produced from this symmetry transformations.

\section{Dyonic static solutions with axial symmetries}
\label{Sec:dyonic_static}

In this section, we consider the more general solution when both electric and magnetic charges present. Similar to the previous section we obtain Maxwell-TEGR and Maxwell-$f(T)$ gravity with axial symmetry. For this reason, we generalise the gauge potential \eqref{eq:gauge_potential} to also include electric charge:
\be {\cal A}= {\cal A}_0(r)\, dt+{\cal A}_4(\theta)\, d\phi.\label{eq:gauge_potential2}\ee

It adds one more non-zero component to the field strength: $F_{rt}=-F_{tr}={\cal A}^{\prime}_0$. The Maxwell equations for this, $r$-dependent, type of components become a little bit more complicated in our metric. More precisely, the magnetic equation remains the same, while the electric one takes the form of
$$0=\frac{1}{\sqrt{-g}}\partial_{\mu}\left(\sqrt{-g}F^{\mu t}\right)\propto
\partial_r\left( \sqrt{W} r^2\cdot F^{rt}\right)= -\partial_r\left( \sqrt{W} r^2\cdot \frac{1}{W}\cdot F_{rt}\right)\propto\left(\frac{r^2}{\sqrt{W}}{\cal A}^{\prime}_0\right)^{\prime}$$
which means $F_{rt}={\cal A}^{\prime}_0=-q\frac{\sqrt{W(r)}}{r^2}$, where the arbitrary constant is chosen as $-q$ so as to correspond to the usual definition of charge with ${\cal A}_0=\frac{q}{r}$ in case of $W=1$.

We see that in case of non-constant $W(r)$, the electric potential behaves not in the standard way which is familiar from the monopole law of $\propto\frac{1}{r}$. This change is dictated by the Maxwell equations in our metric. This important feature was fully ignored in the paper \cite{Capozziello:2012zj}, therefore producing wrong solutions.

We see that, superficially, electric part behaves differently. However, it is very easy to see that the energy-momentum tensor acquires basically the very same form as before:
$$\kappa{\cal T}^{\nu}_{\mu}=\mathrm{diag} \left(-\frac{p^2+q^2}{r^4}, \, -\frac{p^2+q^2}{r^4}, \, \frac{p^2+q^2}{r^4},\, \frac{p^2+q^2}{r^4} \right),$$
which shows the electromagnetic duality in its full power. We present the dyonic solutions below. But it must be clear by now that everything can be obtained from the magnetic case by substituting $p\longrightarrow \sqrt{p^2+q^2}$, and the same way as the sign of $p$ had no relation to the actual sign of the magnetic charge but rather represented a bifurcation of possible solutions, below one also has to bare in mind that this square root can be defined with either plus or minus sign.

\subsection{GR/TEGR}
\label{Sec:dyonic_static_TEGR}

For GR/TEGR, $f(T)=T_0+T$, using the ansatz \eqref{eq:static_tetrad} for the tetrad and the gauge potential \eqref{eq:gauge_potential2} one can show that $A$, $B$, ${\cal A}_0$ and ${\cal A}_4$ satisfying  the field equations are
\be A=C\cdot\left(-{T_0 \over 6}\, r^2 -{2m \over r} + {p^2 +q^2 \over r^2}\right) \ee
\be W= C,  \hspace{0.3 in} {\cal A}_0=\sqrt{C}\cdot {q \over r},\hspace{0.3 in} {\cal A}_4= p\, \theta ,\ee
where $q$ denotes the BH electric charge. Similar to Subsection \ref{Sec:magnetic_static_TEGR} we can set $C=1$ without any change to the solution. 
The torsion scalar \eqref{eq:static_Tsc} of this solution is
\be
T(r)= T_0+\frac{2(q^2+p^2)}{r^4}.\label{eq:dyonic_sing}
\ee
We can illustrate this result by a very simple calculation again. In the radial equation we simply equate the quantity $\frac12 f - f_T T=\frac12 (T_0 -T)$ to the matter source $\kappa{\cal T}^r_r=2\left(F_{rt}F^{rt}-\frac14 \left(2F_{rt}F^{rt}+2F_{\theta\phi}F^{\theta\phi}\right)\right)=-\frac{q^2 + p^2}{r^4}$ and get this answer.

\subsection{Quadratic polynomial $f(T)$ gravity}
\label{Sec:dyonic_static_fT}

Substituting the energy-momentum tensor to the right-hand side of the equations with $f(T)= -\frac{1}{12\beta}+ T + \beta T^2$, we obtain the torsion scalar
\be T = -{1 \over 6 \beta}\pm \sqrt{\frac{2}{3\beta}}\cdot{\sqrt{p^2+q^2 }\over r^2}\ee
and the solution in the form
\be A =C\cdot\left( {r^2 \over 36 \beta}-{2m \over r}+{3\left( p^2+q^2\right) \over 2 r^2}+{\sqrt{6\beta\left( p^2+q^2\right)^3} \over 3 r^4} \right), \quad W=C\cdot \left(1+{\sqrt{6\beta \left(p^2+q^2\right)} \over r^2}\right)^2
\label{potquad} \ee
where we omitted the plus-minus sign assuming an arbitrary sign in the definition of $\sqrt{p^2+q^2}$ as was mentioned above. And, needless to say, we can put $C=1$. In absence of magnetic and electric charges the teleparallel torsion scalar is constant $T_c=2T_0=-\frac{1}{6\beta}$.

Regarding the EM potentials, we have ${\cal A}_4= p \theta$ as before. For the electric part, we know that ${\cal A}^{\prime}_0=-q\frac{\sqrt{W(r)}}{r^2}$. Setting $C=1$ and using the gauge freedom to omit an arbitrary constant additive term, we have
\begin{equation}
{\mathcal A}_0=\frac{q}{r}\cdot \left(1+{\sqrt{6\beta \left(p^2+q^2\right)} \over 3r^2}\right).
\end{equation}
Let us mention again that this behaviour is different from classical electrodynamics due to non-constant nature of $W(r)$, and it was ignored in the paper \cite{Capozziello:2012zj} leading to wrong solutions.

Obviously in absence of the electric charge $q$ the solution reduces to the magnetically charge solution. Also, by setting $q=p=0$, one gets a constant torsion solution. It is worth noticing here that if we try here to generate another solution from this dyonic one, through electric-magnetic duality transformation we end up with the same solution. One can see that by taking $q\rightarrow p$ and $p\rightarrow -q$, it interchanges $q$ and $p$ in the dyonic solution. But since the solution is the same as we interchange $q$ and $p$, no new solution is produced from this symmetry transformations.
In the following we discuss the main physical features which characterize the obtained solution.

\subsection{Physical features of the Black Hole}
\label{Sec:Phy_BH}

\begin{figure}
\centering
\subfigure[~Possible horizons]{\label{fig:1a}\includegraphics[scale=.3]
{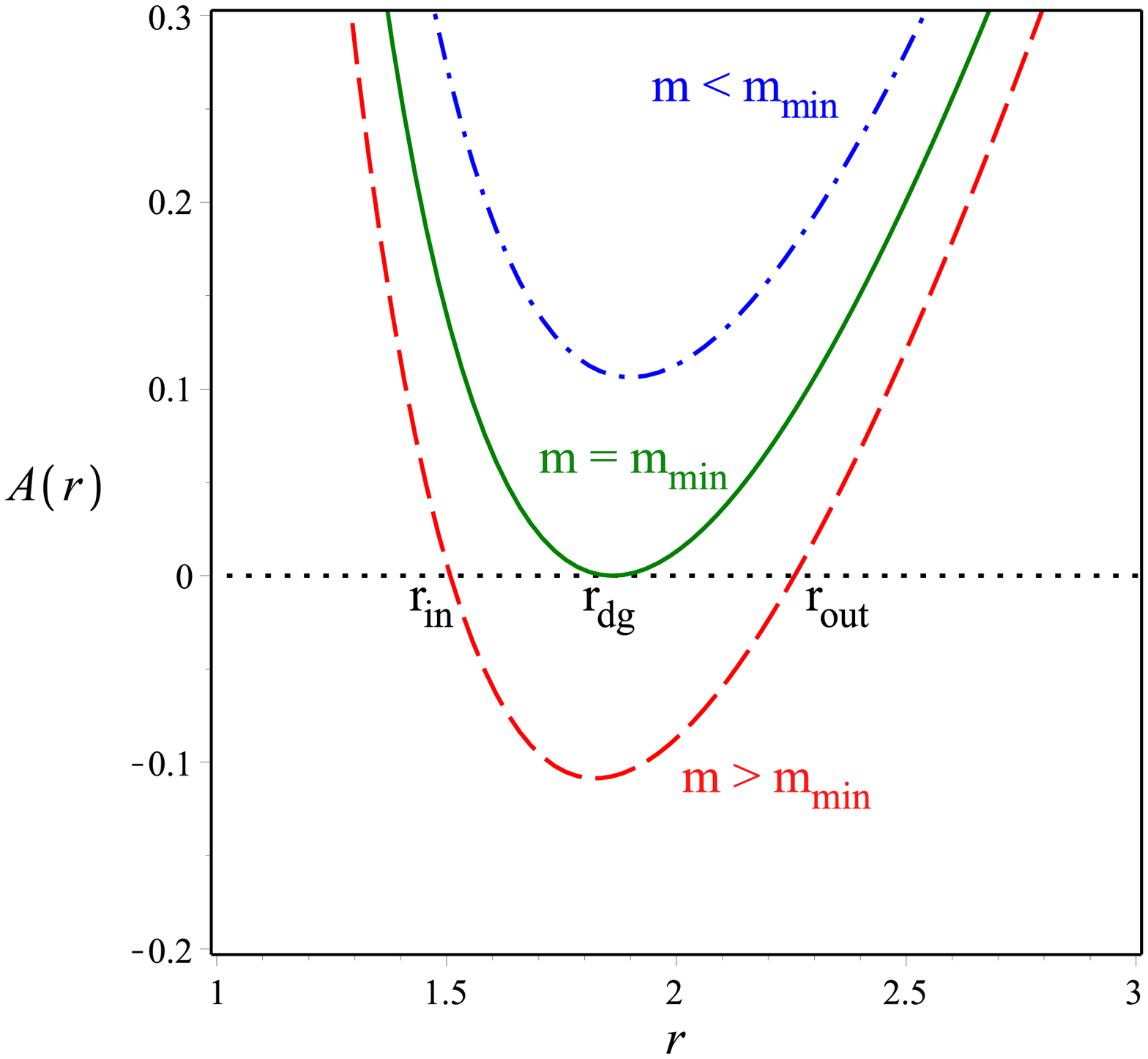}}\hspace{.7cm}
\subfigure[~Horizon mass-radius relation]{\label{fig:1b}\includegraphics[scale=.3]
{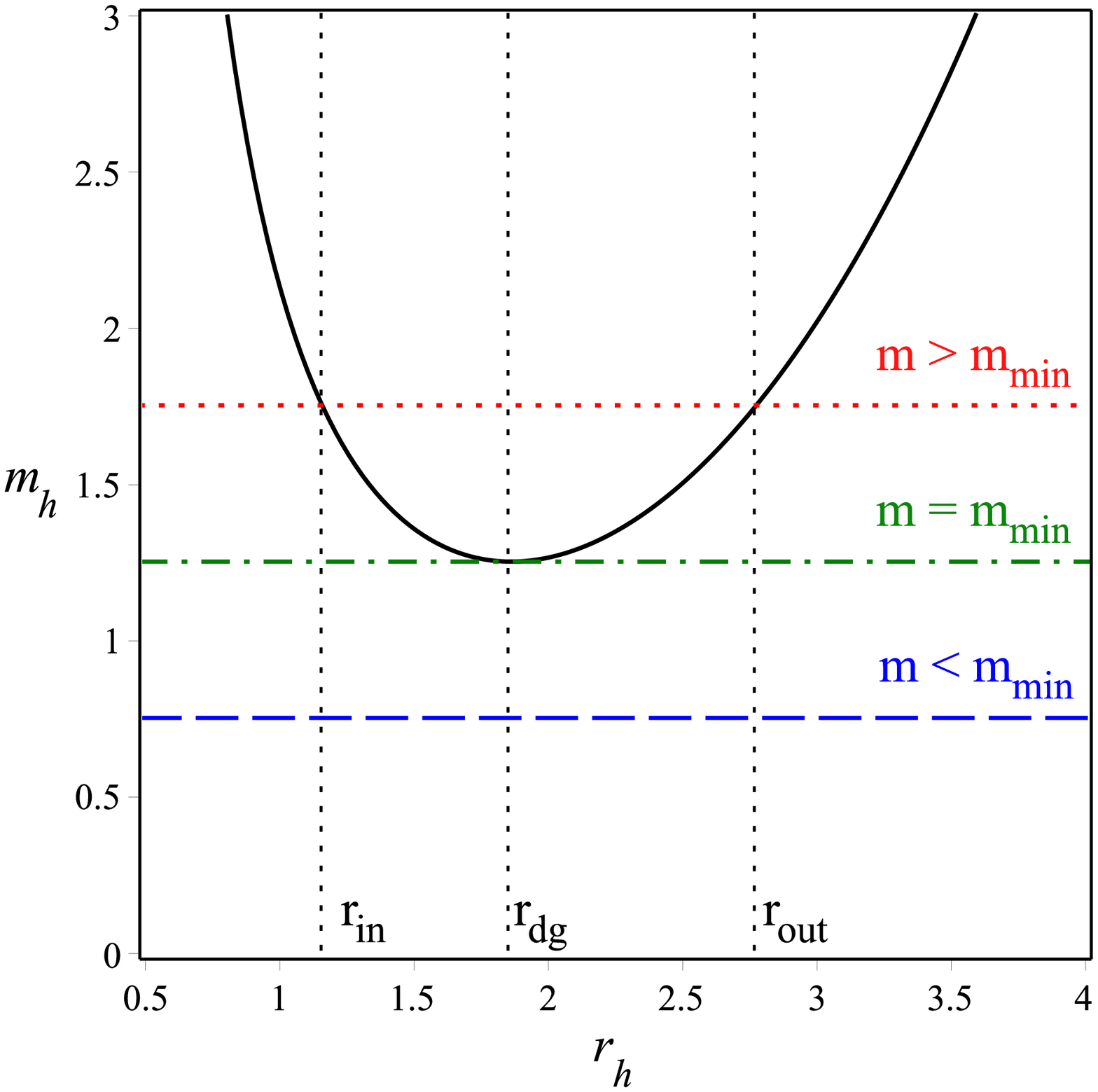}}
\caption[figtopcap]{Schematic plots of possible horizons: \subref{fig:1a}
{\it{Typical behaviour of the function $A(r)$ given by \eqref{potquad}.}}
\subref{fig:1b}
{\it{The horizon mass-radius relation \eqref{eq:quad_hor_m-r} with the minimum mass \eqref{eq:quad_m_min} at which the two horizons coincide at the degenerate horizon \eqref{eq:quad_r_dg}, where $m>m_\text{min}$ the BH has two horizons ($r_{in}$ and $r_{out}$) at most, while for $m<m_{min}$ case the BH is naked. We take $\beta=1/4$ and $q=p=1$.}}}
\label{Fig1}
\end{figure}

BH horizon is one of the most important features which characterizes BHs. We restrict the discussion to the feasible case $\beta>0$. Recalling the metric potentials \eqref{potquad}, we see that the polynomial $A(r)r^4$ has only one term with a negative coefficient. Therefore, by the virtue of Descartes' sign rule, it can have at most two positive real roots. Then  the solution can produce at most two horizons where $A(r)=0$, see Fig. \ref{Fig1}\subref{fig:1a}, the inner, $r_{in}$, and the outer, $r_{out}$, horizons. Note that the function $W(r)$ does not produce new horizons since it is always non-zero positive function. 

Note however that, for the regions of $A(r)<0$, the tetrad fields \eqref{eq:static_tetrad} become complex, though the measurable quantities (metric) and the action stay real. This happens as well for many tetrad solutions with an event horizon, and is unavoidable in our approach. Indeed, we use coordinates in which some components of the diagonal metric flip their signs, and we assume a tetrad which is diagonal in these coordinates. For example, in TEGR or in GR with tetrad fields, one can construct Schwarzschild solution where the tetrad is complex inside the horizon. Since the tetrad is a dynamical variable, in the teleparallel framework, it still can be viewed as a problem from the foundational point of view. The nice point is that our tetrad is always real outside the horizon, i.e. in the potentially observable region. The properties of the horizon and beyond should become a subject of a separate investigation. Even though the metric geometry is perfectly smooth across the horizon, the same can hardly be said about the tetradic structure, for it has a square root of a function whose value passes through zero, and some scalar quantities such as $\accentset{\circ}{\nabla}_{\mu} e^{\mu}_0 $ appear to be singular there. One more interesting point to mention is that the region in between the horizons can also be done with a real tetrad by changing the Minkowski metric there from $\mathrm{diag}( -1, +1, +1, +1)$ to $\mathrm{diag}(+1, -1, +1, +1)$.

When the two roots of $A(r)=0$ merge, this case of the extremal BH is characterized by $A=0$ and $A'=0$. We find that those two horizons are possible as long as the BH mass exceeds a minimal value
\begin{equation}
    m_\text{min}=\frac{7}{36}\sqrt[4]{54 (q^2+p^2)^3 \over \beta}.\label{eq:quad_m_min}
\end{equation}
This defines the minimum horizon mass-radius relation. At $m=m_\text{min}$, the two horizons coincide forming a degenerate horizon ($r_{in}=r_{out}=r_{dg}$), where
\begin{equation}
    r_{dg}=\sqrt[4]{24\beta(q^2+p^2)}.\label{eq:quad_r_dg}
\end{equation}
In this case the solution is called extremal solution. 

For the case $m<m_\text{min}$ the solution possesses a naked singularity, i.e., no BH solution. By setting $A(r_h)=0$, one gets a relation between the mass parameter, $m$ and the horizon radius $r_h$
\begin{equation}
    m_h\equiv m(r_h)={r_h^3 \over 72 \beta}+{3(q^2+p^2) \over 4 r_h}+ {[6\beta(q^2+p^2)]^{3/2} \over 36 \beta r_h^3}.\label{eq:quad_hor_m-r}
\end{equation}
This relation is depicted by Fig. \ref{Fig1}\subref{fig:1b} for certain values of the parameters. Notice that another way of obtaining the degenerate horizon $r_h=r_{dg}$ is to solve $dm_h/dr_h=0$ for $r_h$, which gives the same result as in \eqref{eq:quad_r_dg}. Then, by substituting \eqref{eq:quad_r_dg} into \eqref{eq:quad_hor_m-r}, the extremal mass \eqref{eq:quad_m_min} is re-obtained.\\

\subsubsection{Singularities}
One of the interesting features of the magnetic and dyonic solutions presented here is that they have milder singularities compared to that of constant torsion solutions. Calculating the leading term of the torsion scalar one obtains
\begin{equation}
    T={\sqrt{6\beta (q^2+p^2)} \over 3 \beta r^2} +O(r^{0}),
\end{equation}
which has a milder singularity as $r\rightarrow 0$ compared to the GR/TEGR magnetic solution presented above. Furthermore, Ricci scalar, Ricci tensor square and Kretschmann scalar have milder behavior near $r=0$ too,
\bea
    &&R=-{\sqrt{6(q^2+p^2)/\beta} \over 3 r^2} +O(r^{0}),\,\,\,\, R^{\mu\nu} R_{\mu\nu}={7 (q^2+p^2) \over 9 \beta r^4} +O(r^{-2}),\nonumber\\
    &&R^{\mu\nu\alpha\beta}R_{\mu\nu\alpha\beta}= {10(q^2+p^2) \over 3 \beta r^4} +O(r^{-2}),
\eea
where we allow ourselves the freedom of not putting circles over the Levi-Civita quantities.
To compare with the GR/TEGR solutions, let us list the above invariants as $r\rightarrow 0$
\bea
 && T={2(q^2+p^2) \over r^4}+O(r^{0}), \hspace{1in} R=2\,T_0 ,\nonumber\\ && R^{\mu\nu} R_{\mu\nu}={4(q^2+p^2)^2 \over r^8}+O(r^{0}), \,\,\,\, R^{\mu\nu\alpha\beta}R_{\mu\nu\alpha\beta}= {56(q^2+p^2)^2 \over r^8}+O(r^{-6}).
\eea 
Notice that Ricci scalar is regular in the constant-torsion cases but diverges for our solution.
This shows that singularities within the quadratic polynomial teleparallel gravity could be milder than the corresponding GR solution. But one might ask, can we extend a test particle trajectory beyond the singular point? Is it going to be less singular than that of the GR/TEGR case? Let us check a simple trajectory, namely, a null-radial trajectory for this case. It has the following equation \be {dr \over dt} =\pm {A(r)\over \sqrt{W(r)}}.\ee
In the local region around $r=0$ we have 
\be {dr \over dt} =\pm {(q^2+p^2)\over 3 r^2}+O(r^0).\ee
These trajectories can not be extended beyond the singular point at $r=0$, since the right hand side is neither continues nor differentiable at $r=0$. Therefore, the local solution around $r=0$, is neither guaranteed to exist nor it is unique. Our simple calculation shows that in spite of the milder form of curvature and torsion invariants the null-radial trajectories are still in-extendable beyond the singular point as in the GR/TEGR case.

\subsubsection{Conserved Quantities}

Defining a conserved charge in GR, TEGR or their extensions is always a nontrivial task, especially if we want to link it with the symmetries of the underlying field theory. The authors in \cite{Maluf:2002zc} used a Hamiltonian analysis to find the energy and momentum of any stationary gravitational solution in TEGR gravity. Later this expression was extended to $f(T)$ gravity by the authors in \cite{Ulhoa:2013gca}. Asymptotically, these charges produce the conserved 4-momentum in Minkowski space. This 4-vector, $P^a\equiv(E, \mathbf{P})$ in $f(T)$ gravity \cite{Ulhoa:2013gca} is given by 
\begin{equation}
P^a=\frac{1}{8 \pi}\int_V d^3 x \, \partial_\nu \Pi^{a\nu}=\frac{1}{8 \pi}\oint_\Sigma d\Sigma_\nu \, \Pi^{a\nu},
\label{eq:grav_EM_vec}
\end{equation}
where $\Pi^{a\nu}=e S^{a0\nu} f_T(T)$. The above expression represents the total energy-momentum contained in a three-dimensional volume of space $V$ bounded by the surface $\Sigma$. In order to get the temporal component $P^0$, we evaluate
\begin{equation}
    S^{(0)(0)r}\equiv e{^0}{_\mu}e{^0}{_\nu} S^{\mu \nu r}= -{A \over r W}.
    \label{eq:Supot_temp}
\end{equation}
Then we use the regularized expression of the energy–momentum \eqref{eq:grav_EM_vec}, which guarantees its vanishing for the flat spacetime
\begin{equation}
P^a_{reg}=\frac{1}{8 \pi} \oint_\Sigma d\Sigma_\nu \, [\Pi^{a\nu}-\Pi^{a\nu}_{dS}],
\label{eq:Reg_grav_EM_vec}
\end{equation}
where $\Pi^{a\nu}_{dS}$ denotes the contribution of pure de Sitter spacetime by requiring the physical parameters ($m,\, q,\, p$) to vanish. Substituting \eqref{eq:Supot_temp} into \eqref{eq:Reg_grav_EM_vec}, for the solution set \eqref{potquad}, we get
\begin{equation}
    E_{reg}={m \sigma \over 6 \pi}-{(q^2+p^2) \sigma \over 8 \pi r}- {\sqrt{6\beta}(q^2+p^2)^{3/2}\, \sigma \over 36 \pi r^3}+O\left(\frac{1}{r^5}\right).
\end{equation}
By taking the relation $m=\tilde{G}M$ where $M$ is the BH physical mass, $\tilde{G}=G/f_T(T_c)$; at spatial infinity $r\to \infty$ one obtains the total energy $E={M\sigma \over 4 \pi}$ as measured by a stationary observer at infinity. Here $\sigma$ is the area of the flat surface, or $\sigma=\int d\theta \int d\phi$. This mass or total energy is the same as the one obtained for the flat horizon BH in GR/TEGR case.\\

\section{Conclusions}
\label{sec:conclusions}

BH solutions represent an intricate topic in $f(T)$ gravity theories. The standardly used spherical coordinates do not allow for a diagonal tetrad solution (in pure tetrad gauge) with non-constant torsion scalar, and the non-diagonal "good" tetrads lead to very complicated equations. It prevents many researchers from finding new non-trivial exact BH solutions in this framework.

In this work, we give a detailed review of the issue of diagonal tetrads for the spherically symmetric class of spacetimes, with an overview of the constant $T$ options. Most importantly, we have shown that in the case of flat horizons one can use Cartesian coordinates on the horizon which allow for a simple diagonal tetrad which solves the equations of motion, without requiring a constant torsion scalar. Furthermore, we have generalised this interesting result by proving the following statement: if all the components of a diagonal tetrad depend on only one of the coordinates in which it is written, then this tetrad automatically solves the antisymmetric part of equations. Based on that, we presented non-trivial exact topological BH solutions of magnetic and dyonic type.
Usually the problem with simple tetrad choices in $f(T)$ is in the anti-symmetric equations not being satisfied. However, for BHs with flat horizon, these equations are satisfied if we used Cartesian coordinates on the horizon because of the above Theorem. It is the $rr$ component what makes the equations demand constant $T$. And very interestingly, the addition of matter leads to the possibility of non-constant $T$ solutions. 

By extending our investigation to the time-dependent cases, we find that diagonal tetrads with spherical and hyperbolic symmetries lead to trivial solutions with constant torsion scalar just as the static case. Interestingly, for the time dependent tetrad with a planar symmetry (or a flat horizon), the anti-symmetric part of the field equations becomes non-trivial and generically does not lead to a constant torsion scalar when Cartesian coordinates on the horizon are used. In this case, the presence of matter is required for existence of non-trivial consistent solutions. We leave this case to future work.

The new solutions presented here are characterised by mass, electric and magnetic charges as well as their flat horizons which could be a torus or a cylinder, depending on the global identifications. These solutions possess two horizons, inner and outer ones, that coincide in the extremal case, i.e., $m=m_\text{min}$. Below the extremal mass the solutions form a naked singularity. These solutions possess milder singularities at $r=0$ (compared with GR solutions) as it is clear form the behavior of scalar invariants, apart from Ricci scalar which is divergent in our case. 
It would be interesting to study, in more details, the physical features of these solutions, especially, their geometries and causal structures, as well as the null-like and time-like trajectories in such space-times. It would be interesting to find more of these exact solutions since they are important Laboratories to study the $f(T)$ theories. We hope that we will be able to report on this in our coming works.

\section*{Acknowledgments}

MJG has been supported by the European Regional Development Fund CoE program TK133 “The Dark Side of the Universe” and  by the Estonian Research Council grant MOBJD622.

\end{document}